\PassOptionsToPackage{table}{xcolor}
\documentclass[aps, 10pt,prl,showpacs,twocolumn,longbibliography,superscriptaddress]{revtex4-1}
\usepackage{times}
\usepackage{hyperref}
\usepackage{graphicx}
\usepackage{pgf}
\usepackage{color}
\usepackage{multirow}
\usepackage{mathtools}

\usepackage{pdfpages} 
\usepackage{pgffor} 
\makeatletter
\AtBeginDocument{\let\LS@rot\@undefined}
\makeatother

\begin{document}
\title{Evidence of nodal gap structure in the basal plane of the FeSe superconductor}

\author{Pabitra K. Biswas}
\email[]{pabitra.biswas@stfc.ac.uk}
\affiliation{ISIS Pulsed Neutron and Muon Source, STFC Rutherford Appleton Laboratory, Harwell Campus, Didcot, Oxfordshire, OX11 0QX, United Kingdom}

\author{Andreas Kreisel}
\affiliation{Institut f\"ur Theoretische Physik, Universit\"at Leipzig, D-04103 Leipzig, Germany}

\author{Qisi Wang}
\affiliation{State Key Laboratory of Surface Physics and Department of Physics, Fudan University, Shanghai 200433, China}

\author{Devashibhai T. Adroja}
\affiliation{ISIS Pulsed Neutron and Muon Source, STFC Rutherford Appleton Laboratory, Harwell Campus, Didcot, Oxfordshire, OX11 0QX, United Kingdom}
\affiliation{Highly Correlated Matter Research Group, Physics Department, University of Johannesburg, PO Box 524, Auckland Park 2006, South Africa}

\author{Adrian D. Hillier}
\affiliation{ISIS Pulsed Neutron and Muon Source, STFC Rutherford Appleton Laboratory, Harwell Campus, Didcot, Oxfordshire, OX11 0QX, United Kingdom}

\author{Jun Zhao}
\affiliation{State Key Laboratory of Surface Physics and Department of Physics, Fudan University, Shanghai 200433, China}

\author{Rustem Khasanov}

\author{Jean-Christophe Orain}

\author{Alex Amato}

\author{Elvezio Morenzoni}
\affiliation{Laboratory for Muon Spin Spectroscopy, Paul Scherrer Institut, CH-5232 Villigen PSI, Switzerland}

\date{October 11, 2018}




\begin{abstract}
Identifying the symmetry of the wave function describing the Cooper pairs is pivotal in understanding the origin of high-temperature superconductivity in iron-based superconductors. Despite nearly a decade of intense investigation, the answer to this question remains elusive. Here we use the muon spin rotation/relaxation ($\mu$SR) technique to investigate the underlying symmetry of the pairing state of the FeSe superconductor, the basic building block of all iron-chalcogenide superconductors. Contrary to earlier $\mu$SR studies on powders and crystals, we show that while the superconducting gap is most probably anisotropic but nodeless along the crystallographic $c$-axis, it is nodal in the $ab$-plane, as indicated by the linear increase of the superfluid density at low temperature. We further show that the superconducting properties of FeSe display a less pronounced anisotropy than expected.
\end{abstract}

\maketitle

High transition-temperature $T_{\rm c}$ superconductivity in Fe-based materials is an intriguing emergent phenomena in modern condensed matter physics research \cite{Paglione2010, Stewart2011, Wang2011, Wen2011, Chubukov2015}. Among various Fe-based superconductors, FeSe is one of the most interesting and intensively studied compounds due to its extremely simple crystal structure, high $T_{\rm c}$ values, unconventional superconducting state and unusual normal state properties. Superconductivity takes place in the FeSe layer which is the basic building block of all Fe-chalcogenide superconductors\cite{Bohmer_review}.
Despite nearly a decade of extensive research, the symmetry of the superconducting gaps in FeSe, which is intimately connected to the electrons pairing mechanism in this material and all other related Fe-based superconductors, is still subject of intense debate. While anisotropic line nodes or deep minima in the superconducting gaps have been suggested theoretically in FeSe~\cite{Mukherjee2015}, most experimental techniques have detected two superconducting gaps, however without any consensus about the presence or absence of nodes in either of the gaps~\cite{Khasanov2008, Dong2009, Khasanov2010, Ponomarev2011, Lin2011, Hafiez2013, Kasahara2014, Bourgeois-Hope2016,Watashige17}. Notable exceptions are surface sensitive scanning tunnelling spectroscopic (STS) measurements performed on FeSe thin films, which detected “V”-shaped conducting spectra in the superconducting state, indicating the presence of nodes in the gap structure~\cite{Song2011}. A similar STS experiment conducted on the twin boundaries of FeSe single crystals displayed a fully gapped structure, suggesting a gap-symmetry evolution from nodal in the bulk to nodeless at the twin boundaries~\cite{Watashige2015}, a finding that has been argued to be in agreement with the detection of a finite gap in multiple domains while in single domains the gap is found to be zero within experimental resolution\cite{Hashimoto18}. Recently, Sprau \textit{et al.} used a quasiparticle interference imaging technique and detected gap minima in the $\alpha$ and $\epsilon$ bands of the Fe plane~\cite{Sprau2017}. They further suggested that the Cooper pairing in FeSe is orbital-selective, involving predominantly the $d_{\rm yz}$ orbitals of the Fe atoms. However, the majority of the techniques used so far in detecting nodes or gap-minima are surface sensitive only and give limited or no information about the symmetry of the pairing state in the bulk of FeSe. To date, there is no clear and direct bulk evidence of nodes in the gap structure of FeSe. Clarifying this issue is highly desirable not only to determine the exact nature of the superconducting state in FeSe but also because a comparison with the other Fe based superconductors and the cuprates may pave the way to understand the essential ingredients of high-temperature superconductivity.

In this work, we have used the $\mu$SR technique to reveal the symmetry of the superconducting gap along the crystallographic \textit{c}-axis and \textit{ab}-plane of FeSe single crystals. The measurement of the field distribution in the vortex state by $\mu$SR is one of the most direct and accurate methods to determine the absolute value of the magnetic penetration depth $\lambda$ and its temperature dependence \cite{Sonier2000}. $\lambda(T)$ is related to the effective superfluid density, the density of the superconducting carriers $n_s$ as $\lambda^{-2}(T) \propto \frac{n_s(T)}{m^*}$, where $m^*$ is the effective mass. The low-temperature behavior of $\lambda(T)$ directly reflects the low-energy properties of the quasi-particle spectrum, and is therefore sensitive to the presence or absence of nodes in the superconducting gap. While for a fully gapped $s$ wave superconductor $\lambda^{-2}(T)$ saturates exponentially with decreasing temperature, it increases linearly in a nodal superconductor \cite{Sonier2000}. Here, we report the direct observation of nodal superconductivity in the basal plane of FeSe. We show that while the temperature dependence of the superfluid density along the crystallographic $c$-axis is compatible with either a nodeless anisotropic $s$ wave or isotropic two-gap $s+s$ wave symmetries, that in the basal $ab$-plane is better fitted assuming a two-gap $s+d$ wave symmetry. The nodal $d$ wave component reflects the linear increase of the superfluid density with decreasing temperature close to $T=0$. The transition of the  pairing symmetry from nodeless to nodal, as we probe from the \textit{out-of-plane} to the \textit{in-plane} direction in the FeSe-layer, suggests a directional dependent pairing symmetry in FeSe.
\\

\begin{figure}[tb]
\includegraphics[width=1.0\linewidth]{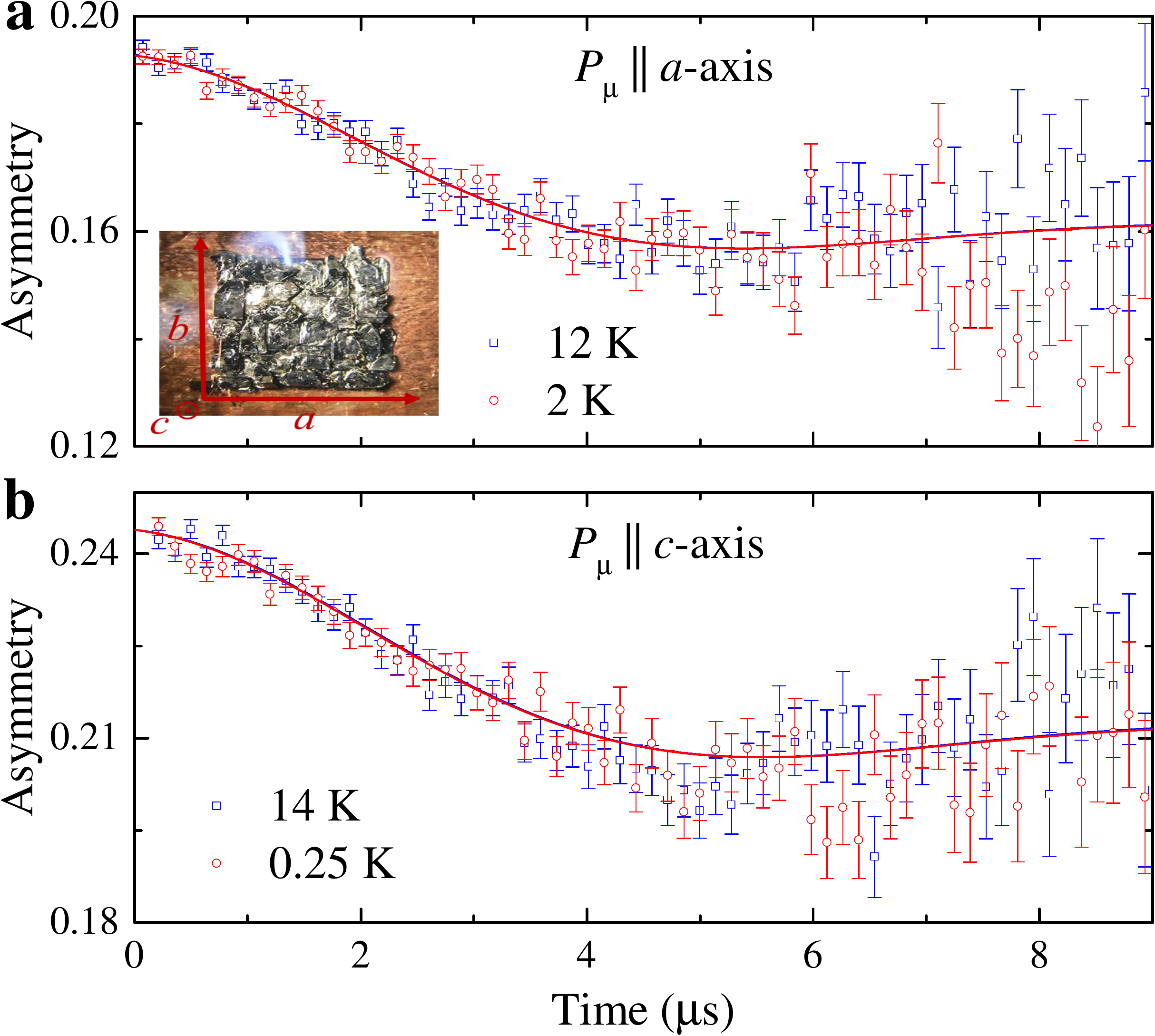}
\caption{ZF-$\mu$SR time spectra, collected above and below $T_{\rm c}$ with muon spin polarization $P_{\rm \mu}$ parallel to the \textbf{a} \textit{a}-axis and \textbf{b} \textit{c}-axis. The solid lines are the fits to the data using the Kubo-Toyabe Gaussian distribution function, described in the text. Inset in \textbf{a} shows the mosaic of the aligned FeSe crystals used in this study.}
 \label{fig:Asy_ZF}
\end{figure}

The sample used in these experiments was an 1 cm$^2$ mosaic of around 30 single crystals, all of them carefully aligned along the three nominal crystallographic axes \textit{a}, \textit{b} and \textit{c}. Details about the crystal growth are described in Ref.~\cite{Wang2015GS}. The crystals were mounted on a 50 $\mu$m thin copper foil, attached to a fork shaped copper sample holder, see Fig.~\ref{fig:Asy_ZF} \textbf{a} inset. Zero-field (ZF) and transverse-field (TF) $\mu$SR experiments were carried out using co-aligned crystals. Figure~\ref{fig:Asy_ZF} \textbf{a} and \textbf{b} show the typical ZF-$\mu$SR time spectra collected above and below $T_{\rm c}$ with muon spin polarization $P_{\rm \mu}$ parallel to the crystallographic \textit{a}- and \textit{b}-axis. The solid lines are the fits to the data using the Kubo-Toyabe Gaussian distribution function, which describes the temporal evolution of the spin polarization in the presence of randomly oriented nuclear moments~\cite{Kubo1981}. Details are described in the Supplemental Materials (SM)\cite{SM}. ZF data collected above and below $T_{\rm c}$ in both orientations do not show any detectable additional relaxation in the asymmetry spectra, therefore completely ruling out the presence of any magnetism in the superconducting state of FeSe.

\begin{figure}[tb]
\includegraphics[width=1.0\linewidth]{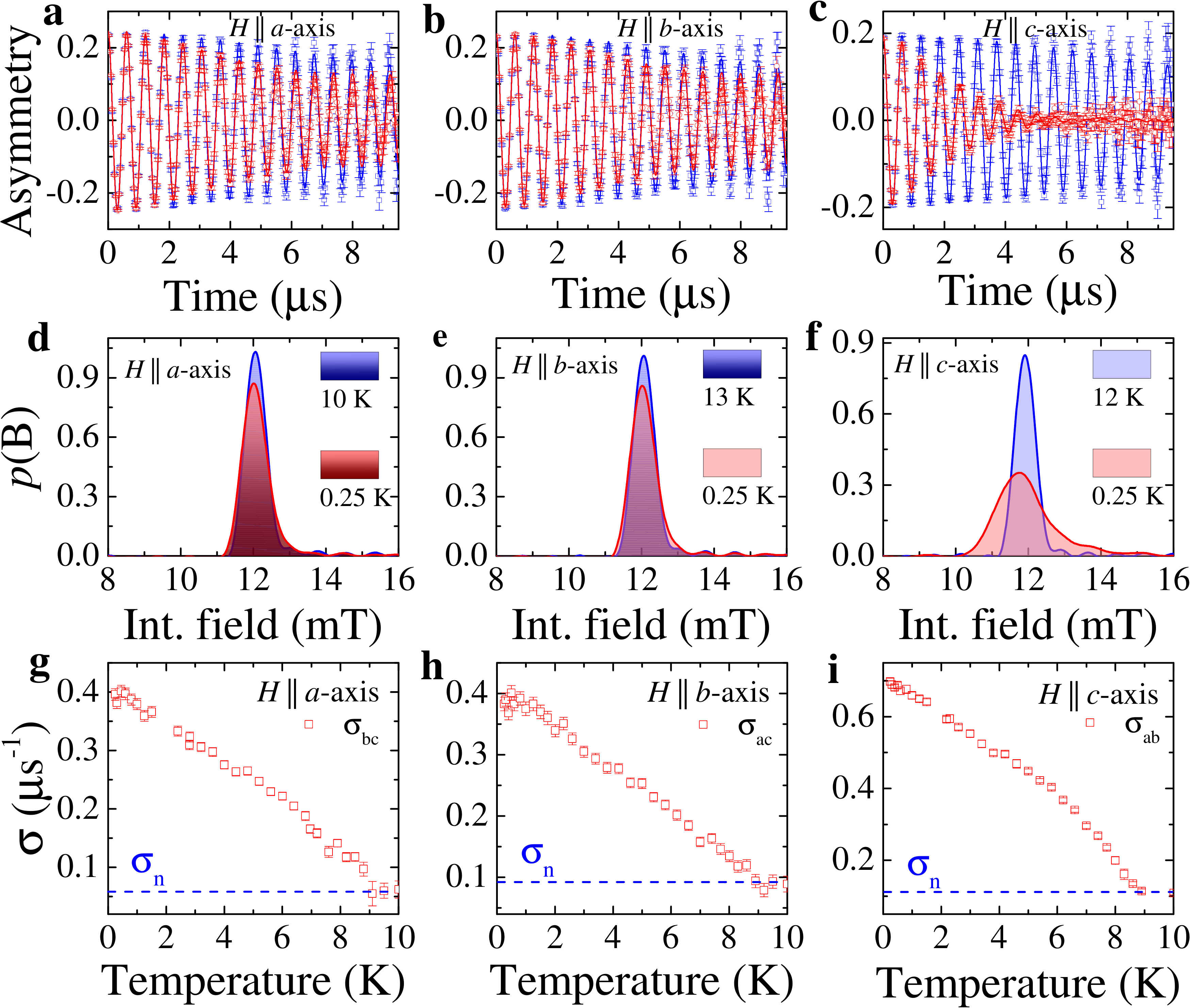}
\caption{\textbf{a}, \textbf{b} and \textbf{c} TF-$\mu$SR time spectra of FeSe, collected above and below $T_{\rm c}$ in a TF of 12 mT applied parallel to the \textit{a}-, \textit{b}- and \textit{c}-axis, respectively. The solid lines are the fit to the data using a sum of Gaussian field distributions, described in the text. \textbf{d}, \textbf{e} and \textbf{f} Fast Fourier transformation of the TF-$\mu$SR spectra, showing the line shape of the internal fields along all three crystallographic axes. \textbf{g}, \textbf{h} and \textbf{i} Temperature dependence of the muon spin damping rate $\sigma$ along three crystallographic directions, extracted from the TF-$\mu$SR time spectra. The dashed horizontal lines represent the normal state contribution $\sigma_{\mathrm{n}}$.}
 \label{fig:Asy_TF_sigma}
\end{figure}

Three sets of TF-$\mu$SR experiments were performed with the magnetic field $H$ applied parallel to three crystallographic axes. Figure~\ref{fig:Asy_TF_sigma} \textbf{a}, \textbf{b} and \textbf{c} show the TF-$\mu$SR asymmetry spectra collected above and below $T_{\rm c}$ with $H=12$ mT applied along the nominal \textit{a}-, \textit{b}- and \textit{c}-axis, respectively. As expected, the TF-$\mu$SR signals decay much faster in the superconducting state than in the normal state  due to the formation of a vortex lattice and the associated inhomogeneous magnetic field distribution. Figure~\ref{fig:Asy_TF_sigma} \textbf{d}, \textbf{e} and \textbf{f} show the fast Fourier transformation (FFT) of the TF-$\mu$SR spectra, revealing the line shape of the internal magnetic field distributions $p(B)$ probed by the muons. Both TF-$\mu$SR time spectra and corresponding FFT clearly demonstrate that the $\mu$SR responses are identical for $H$ applied parallel to the nominal \textit{a}- and \textit{b}-axis. This is expected due to the formation of structural twin domains in FeSe crystals. The background signal is relatively large for $H$ applied parallel to the \textit{a}- and \textit{b}-axis. This is due to the bending of the muon beam under transverse magnetic field to the muon momentum. The field distribution in the FFT signals shows that $p(B)$ is much more asymmetric for $H\parallel{c}$-axis than $H\parallel{a/b}$-axis. Also the damping of the TF-$\mu$SR signals in the superconducting state is much stronger for $H\parallel{c}$-axis than $H\parallel{a/b}$-axis. 

The muon spin depolarization rate $\sigma$ can be determined by fitting the TF-$\mu$SR asymmetry spectra collected with $H\parallel{a/b}$-axis using damped spin precession functions
\begin{align}
\label{Depolarization_Fit}
A_{TF}(t)=&A_0\exp\left(-\sigma^{2}t^{2}\right/2)\cos\left(\gamma_\mu \left\langle B\right\rangle t +\phi\right)\notag\\
&
+A_{\rm bg}\cos\left(\gamma_\mu B_{\rm bg}t +\phi\right),
\end{align}
where $A_0$ and $A_{\rm bg}$ are the initial asymmetries of the sample and background signals, respectively, $\gamma_{\mu}/2\pi=135.5$~MHz/T is the muon gyromagnetic ratio~\cite{Sonier2000}, $\left\langle B\right\rangle$ and $B_{\rm bg}$ are the internal and background magnetic fields, and $\phi$ is the initial phase of the muon precession signal. In order to account for the highly asymmetric nature of $p(B)$, TF-$\mu$SR asymmetry spectra collected for $H$ applied parallel to the \textit{c}-axis were analyzed using the skewed Gaussian (SKG) field distribution, as described in Ref.~\onlinecite{Khasanov2016} (also see SM).

Figure~\ref{fig:Asy_TF_sigma} \textbf{g}, \textbf{h} and \textbf{i} show the temperature dependence of $\sigma$ along all three crystallographic directions, extracted from the TF-$\mu$SR time spectra. 
The depolarization rate can be expressed as the geometric mean of the superconducting contribution to the relaxation rate due to the inhomogeneous field distributions of the vortex lattice, $\sigma_{\mathrm{sc}}$, and the temperature independent nuclear magnetic dipolar contribution $\sigma _{\mathrm{nm}}$, i.e. $\sigma=\sqrt{\sigma^{2}_{\rm sc} + \sigma^{2}_{\rm nm}}$.

\begin{figure}[tb]
\includegraphics[width=1.0\linewidth]{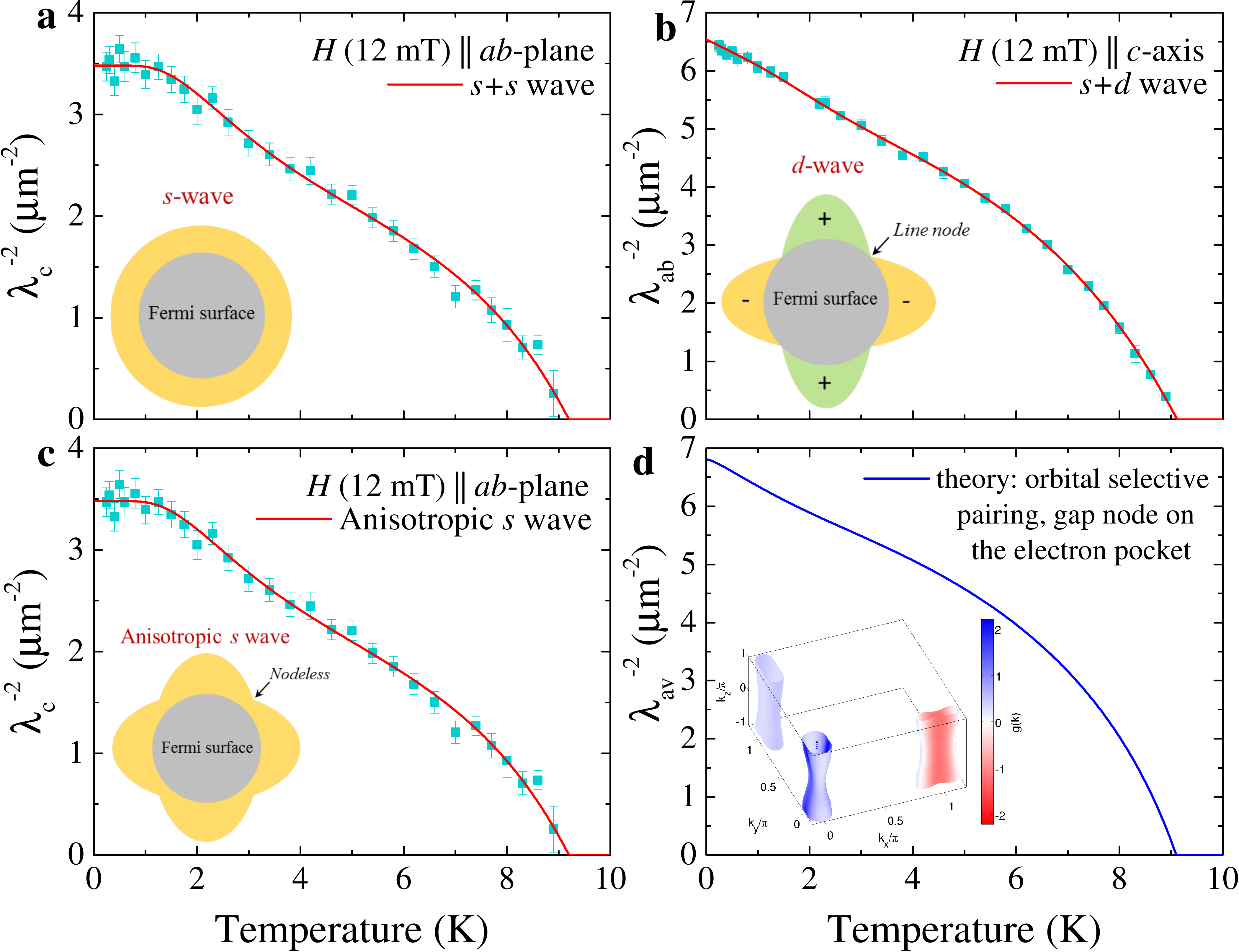}
\caption{\textbf{a}, \textbf{c} Temperature dependence of $\lambda^{-2}$ for FeSe along the crystallographic $c$-axis. The solid curves are the fit to the $\lambda_{\rm c}^{-2}(T)$ using a nodeless anisotropic $s$ wave and two-gap $s+s$ wave models. \textbf{b} Temperature dependence of $\lambda^{-2}$ for FeSe in the $ab$-plane. The solid curves is the fit to the $\lambda_{\rm ab}^{-2}(T)$ using a two-gap $s+d$ wave model.
\textbf{d} Calculation of the averaged penetration depth $\lambda_{\text{av}}^{-2}$ from a microscopic model for the order parameter and the electronic structure.
Insets are the schematic of the isotropic $s$, anisotropic $s$ and $d$ wave type gap symmetries around the Fermi surface and the gap structure for the microscopic model in \textbf{d}.}
 \label{fig:lambda}
\end{figure}

The temperature dependence of the \textit{in-plane} and \textit{out-of-plane} components of the magnetic penetration depth $\lambda_{\rm ab}$ and $\lambda_{\rm c}$ were calculated from $\sigma_{\rm sc}^{\parallel a}$, $\sigma_{\rm sc}^{\parallel b}$ and $\sigma_{\rm sc}^{\parallel c}$ by using the simplified Brandt equation \cite{Brandt1988,Khasanov2016}, as described in Ref.~\onlinecite{Khasanov2016} (also see SM). Figure~\ref{fig:lambda} \textbf{a}, \textbf{b} and \textbf{c} show the temperature dependence of $\lambda^{-2}$ for FeSe along the crystallographic $c$-axis and $ab$-plane, respectively. The solid curves are the fit to the $\lambda^{-2}(T)$ using either a single-gap or a two-gap model,
\begin{equation}
\label{two_gap}
\frac{\lambda^{-2}\left(T\right)}{\lambda^{-2}\left(0\right)}=\omega\frac{\lambda^{-2}\left(T, \Delta_{0,1}\right)}{\lambda^{-2}\left(0, \Delta_{0,1}\right)}+(1-\omega)\frac{\lambda^{-2}\left(T, \Delta_{0,2}\right)}{\lambda^{-2}\left(0, \Delta_{0,2}\right)}.
\end{equation}
Here $\lambda\left(0\right)$ is the value of the penetration depth at $T=0$~K, $\Delta_{0,i}$ is the value of the $i$-th ($i=1$ or 2) superconducting gap at $T=0$~K and $\omega$ is the weighting factor of the first gap. Each term in Eq. (\ref{two_gap}) is evaluated using the standard expression within the local London approximation ($\lambda \gg \xi$)~\cite{Prozorov2006} as
\begin{equation}
\frac{\lambda^{-2}\left(T, \Delta_{0,i}\right)}{\lambda^{-2}\left(0, \Delta_{0,i}\right)}=1+\frac{1}{\pi}\int^{2\pi}_{0}\!\int^{\infty}_{\Delta_{\left(T,\varphi\right)}}\!\frac{\left(\frac{\partial f}{\partial E}\right) EdE d\varphi}{\sqrt{E^2-\Delta_i^2\!\left(T,\varphi\right)}},
\end{equation}
where $f=[1+\exp(E/k_{\rm B}T)]^{-1}$ is the Fermi function, $\varphi$ is the angle along the Fermi surface, and \mbox{$\Delta_i\left(T,\varphi\right)=\Delta_{0, i}\delta\left(T/T_c\right)g\left(\varphi\right)$}, where $g\left(\varphi\right)$ describes the angular dependence of the gap. $g\left(\varphi\right)$ is 1 for $s$ wave and $s+s$ wave gaps, $\left|\cos\left(2\varphi\right)\right|$ for a $d$ wave and $\left[1+a\left|\cos\left(4\varphi\right)\right|\right]$ for anisotropic $s$ wave gap. An approximation to the temperature dependence in $\Delta(T)$ can be written as \mbox{$\delta\left(T/T_{\rm c}\right)=\tanh\bigl\{1.82\left[1.018\left(T_{\rm c}/T-1\right)\right]^{0.51}\bigr\}$~\cite{Carrington2003}}.

All the fitted parameters are summarized in Table~\ref{table} and details about the fit functions are described in SM. For the superfluid density along the $c$-axis, i.e. $1/\lambda_{\rm c}^2\left(T\right)$ ($H||ab$-plane), both the single-gap anisotropic $s$ wave and two-gap $s+s$ wave gap models give the lowest $\chi_{\rm reduced}^2$ value and hence represent the best fit to the data compared to any other models tried here. Gap parameters extracted from analysis are in excellent agreement with most of the reported values obtained on this system~\cite{Khasanov2008, Dong2009, Khasanov2010, Ponomarev2011, Lin2011, Hafiez2013, Kasahara2014, Bourgeois-Hope2016,Hashimoto18}.

\definecolor{skyblue}{rgb}{0.53, 0.81, 0.92}
\definecolor{lightblue}{rgb}{0.68, 0.85, 0.9}
\begin{table}[tb]
\caption{Fitted parameters to the $\lambda_{\rm ab}^{-2}(T)$ and $\lambda_{\rm c}^{-2}(T)$ data of FeSe using the different models described in the text.\label{table}}
\begin{tabular}[t]{lllll}\hline\hline
{Data} & Model & Gap value (meV) & $\lambda(0) (\mathrm{nm})$ & $\chi^2_{\rm reduced}$\\\hline
& $s$ wave & $\Delta$=1.22(1) &  & 39.56\\
& $d$ wave & $\Delta$=1.99(2) &  & 4.39\\
& Anisotropic & $\Delta$=1.40(2), $a$=0.81(2) &  & 1.48\\
& $s$ wave & with $\Delta_{\rm Max}$=2.53(4) &  & \\
& $s+s$ wave & $\Delta_1$=1.75(6), $\Delta_2$=0.40(3) &  & 1.40\\
&  & and $\omega$=0.68(6) &  & \\
\multirow{-7}{*}{\cellcolor{white}$\frac{1}{\lambda_{\rm ab}^2(T)}$} & $s+d$ wave & $\Delta_1$=1.86(8), $\Delta_2$=0.73(8) & 391(16) & 1.01\\
 &  & and $\omega$=0.60(2) &  & \\
\hline
& $s$ wave & $\Delta$=1.19(3) &  & 8.41\\
& $d$ wave & $\Delta$=1.9(1) &  & 4.35\\
& Anisotropic & $\Delta$=1.28(4), $a=0.76(4)$ & 514(53) & 1.50\\
& $s$ wave & with $\Delta_{\rm Max}$=2.3(1) &  & \\
& $s+s$ wave & $\Delta_1$=2.2(3), $\Delta_2$=0.6(1) &  & 1.51\\
&  & and $\omega$=0.48(7) &  & \\
\multirow{-7}{*}{\cellcolor{white}$\frac{1}{\lambda_{\rm c}^2(T)}$} & $s+d$ wave & $\Delta_1$=1.8(1), $\Delta_2$=1.0(1) &  & 1.81\\
&  &  and $\omega$=0.44(1) &  & \\\hline\hline
\end{tabular}
\end{table}

For the superfluid density in the $ab$-plane, i.e. $1/\lambda_{\rm ab}^2\left(T\right)$ ($H||c$-axis) we need to introduce a nodal $d$ wave gap along with an isotropic $s$ wave gap in order to reproduce the linear increase of the superfluid density close to zero temperature. We find that the $s+d$ wave model gives a much lower $\chi_{\rm reduced}^2$ value than others. Our results strongly suggest that FeSe is indeed a multigap superconductor.
The experimentally obtained superfluid density in the basal plane shows properties of a nodal superconductor irrespective of the field direction. These findings differ qualitatively from earlier reports on the $\mu$SR studies of FeSe evidencing nodeless superconductivity in this material~\cite{Khasanov2008, Khasanov2010}. This is probably due to the use of polycrystalline samples which is expected to give an average effect from all three directions. It is also well known that the presence of impurities can sometimes mask the true nature of the superconducting gap~\cite{Sonier1999}. Our results are also consistent with the STS measurements performed on FeSe thin films showing nodes in the gap structure~\cite{Song2011}. Recent specific heat data collected on the single crystals of FeSe show a linear behavior at low temperature, a signature that has been interpreted as nodal superconductivity\cite{Jiao_2017,Hardy2018}.  More recently, Y. Sun, \textit{et al.} has performed field-angle-resolved specific heat measurements of FeSe and found three superconducting gaps in FeSe with line nodes in the smaller gap \cite{Sun_2017arXiv}. A strongly anisotropic gap structure with deep minima has been observed in recent quasiparticle interference (QPI) imaging measurements by Sprau \textit{et al.} \cite{Sprau2017}. Anisotropic gap structure has also been found along all momentum directions in a recent ARPES measurements by Kushnirenko \textit{et al.} \cite{Kushnirenko2018}. It is important to note here that both QPI imaging and ARPES are surface sensitive techniques and the deep minima observed at the surface may become node in the bulk of the FeSe superconductor.
\\

To draw conclusions from the measured in-plane and out-of-plane penetration depths beyond the general statement of presence or absence of nodal behavior in certain directions, we also present microscopic calculations of the penetration depth. For this purpose, we start from a recently proposed model for the electronic structure with the eigenenergies $\tilde E_\mu({\mathbf k})$ that is consistent with a number of experimental investigations on FeSe\cite{Kreisel2017,Sprau2017}. The superconducting gap function has been slightly modified to introduce a nodal structure in the bulk of FeSe. Taking into account the electronic structure as being a correlated electron gas via a reduced quasiparticle weight, one can calculate the penetration depth (tensor) without any free parameters. The key ingredient is the parametrization of the Green's function for band $\nu$ in presence of correlations via
$\tilde G_\nu({\mathbf k},\omega_n)=\tilde Z_\nu({\mathbf k})[i \omega_n - \tilde E_\mu({\mathbf k})]^{-1}$
where  $\tilde Z_\nu({\mathbf k})= [\sum_s |a_\nu^s({\mathbf k})|^2 \sqrt{Z_s}]^2$ is the momentum-dependent quasiparticle weight that is obtained from the quasiparticle weights of the orbitals $Z_s$ and the matrix elements $a_\nu^s({\mathbf k})$ for the orbital to band transformation\cite{Kreisel2017,Sprau2017}. The structure of the matrix elements and the values of the quasiparticle weights have been deduced earlier\cite{Kreisel2017,Sprau2017}. Details on the calculation of the inverse square of penetration depth $\lambda_{i}^{-2}$  for shielding supercurrent flowing in $i$ direction are presented in the Supplemental Materials\cite{SM}. At the moment, we simply ignore the contribution of one of the Fermi surface pockets ($\delta$ pocket) to the penetration depth.
In line with the previous theoretical considerations and also in accordance to the expectations of the principal axis of the superfluid tensor\cite{Song2011}, we choose the direction of the short Fe-Fe bond, the long Fe-Fe bond and the crystallographic $c$ axis as directions of our calculations. Noting that the relative magnitudes of $\lambda_x$ and $\lambda_y$ agree to the observed orientation of elongated vortices in FeSe (see SM), we need to keep in mind that the present experiment does not see the difference between the two directions because of the twinning of the crystals.
The geometric mean of the penetration depth in the plane $\lambda_{\text{av}}$ is equivalent to the measured averaged penetration depth $\lambda_{ab}$ due to the tensor nature of the superfluid density\cite{Thiemann1989}, see SM.
In Fig. \ref{fig:lambda} {\bf d} we show the result for $\lambda_{\text{av}}$ from this calculation.
From a theoretical point of view, the full gap is not robust against nodes formation, 
because FeSe in the nematic state allows spherical harmonics from s-wave type gap functions to superimpose to contributions of d-wave symmetry, thus the relative strength of these contributions determines on whether the order parameter goes to zero on the Fermi surface.
The properties of the pairing interaction and thus the superconducting gap can be slightly modified on the surface. Thus our result does not contradict the experimental findings by QPI \cite{Sprau2017}.
Therefore, we used a gap function exhibiting nodes on the electron pocket, see Fig. \ref{fig:lambda}\textbf{d}, inset.
It is evident that the mentioned fully gapped state yields a saturating superfluid density at low temperatures, while the nodal state produces linear behavior in that quantity. A direct comparison of the calculated and measured penetration depth $\lambda^{-2}$ over the full temperature range reveals 
only a difference of 5\% from the experimentally deduced value, an error that can easily be explained by errors in the gap magnitude and the Fermi velocities, see SM.

Table~\ref{table} shows the absolute values of the penetration depth $\lambda(0)$ in both directions. In the basal plane $\lambda(0)$ is 391(16) nm which is lower than the value 514(53) nm  out of the basal plane and reflects the anisotropic superconducting properties in FeSe. Theoretically, a much larger value of $\lambda_c$ is expected given the small dispersion of the proposed electronic structure and the small Fermi velocities in $k_{\rm z}$ direction. Even taking into account a possible misalignment of the external field our results indicate a more 3-dimensional electronic structure for FeSe. From our determination of $\lambda(0)$ and using the reported value of effective mass $m^* \approx 4m_e$~\cite{Watson2015} in the expression for the density of paired electrons $n_{\rm s}(0) = \frac{m^*}{\mu_0 e^2 \lambda^{2}(0)}$, we estimate $n_{\rm s}^{\parallel ab}(0) \approx 7.4\times10^{20}$ cm$^{-3}$ and $n_{\rm s}^{\parallel c}(0) \approx 3.9\times10^{20}$ cm$^{-3}$. These values show that the overall carrier density in FeSe is small, with the basal plane of FeSe playing a preferred role in carrying superconductivity. 
\\

The observation of line nodes in the basal plane of FeSe superconductor is the main finding of our paper. This conclusion does not require a specific theoretical model, but is directly related to the observed low temperature behavior of $1/\lambda^2(T)$, which shows saturation in the out-of-plane and linear increase in the basal plane as the temperature decreases to absolute zero. Such a linear increase of superfluid density reflects the presence of low-energy excitations and thus confirms nodes in the superconducting gap structure of FeSe. To the best of our knowledge this is the first direct experimental demonstration of the existence of nodes in the superconducting gap structure of FeSe using a microscopic bulk probe. These finding offers new insights into the still mysterious superconducting mechanism in iron-based superconductors and may be pivotal to obtain a general understanding of the mechanism of superconductivity among high-$T_{\rm c}$ iron-based and cuprate superconductors.


\textit{Acknowledgments}

D.T.A. would like to thank the Royal Society of London for UK-China Newton funding. Q. W. and J. Z. were supported by the National Natural Science Foundation of China (Grant No. 11374059), the National Key R\&D Program of the MOST of China (Grant No. 2016YFA0300203) and the Ministry of Science and Technology of China (Program 973: 2015CB921302).
Data is available from the corresponding author upon request.

\nocite{Yaouanc}
\nocite{Sonier2000,Brandt1988}
\nocite{Suter2012}
\nocite{Kubo1981}
\nocite{Sonier2000}
\nocite{Maisuradze2009}
\nocite{Brandt1988}
\nocite{Thiemann1989}
\nocite{Mukherjee2015,Kreisel2017,Sprau2017}
\nocite{Kreisel2017}
\nocite{Sheehy04,Eremin10}
\nocite{Eremin10}
\nocite{Sprau2017}
\nocite{Kreisel2017}
\nocite{Rhodes2018,Benfatto2018}
\nocite{Sprau2017,Kreisel2017}
\nocite{Suzuki15,Watson2015,Watsonshibauchi2015PRL_FeSe-QO,Watson2016}
\nocite{Kogan81,Campbell88}
\nocite{Campbell88}

%

\def\supplementfilename{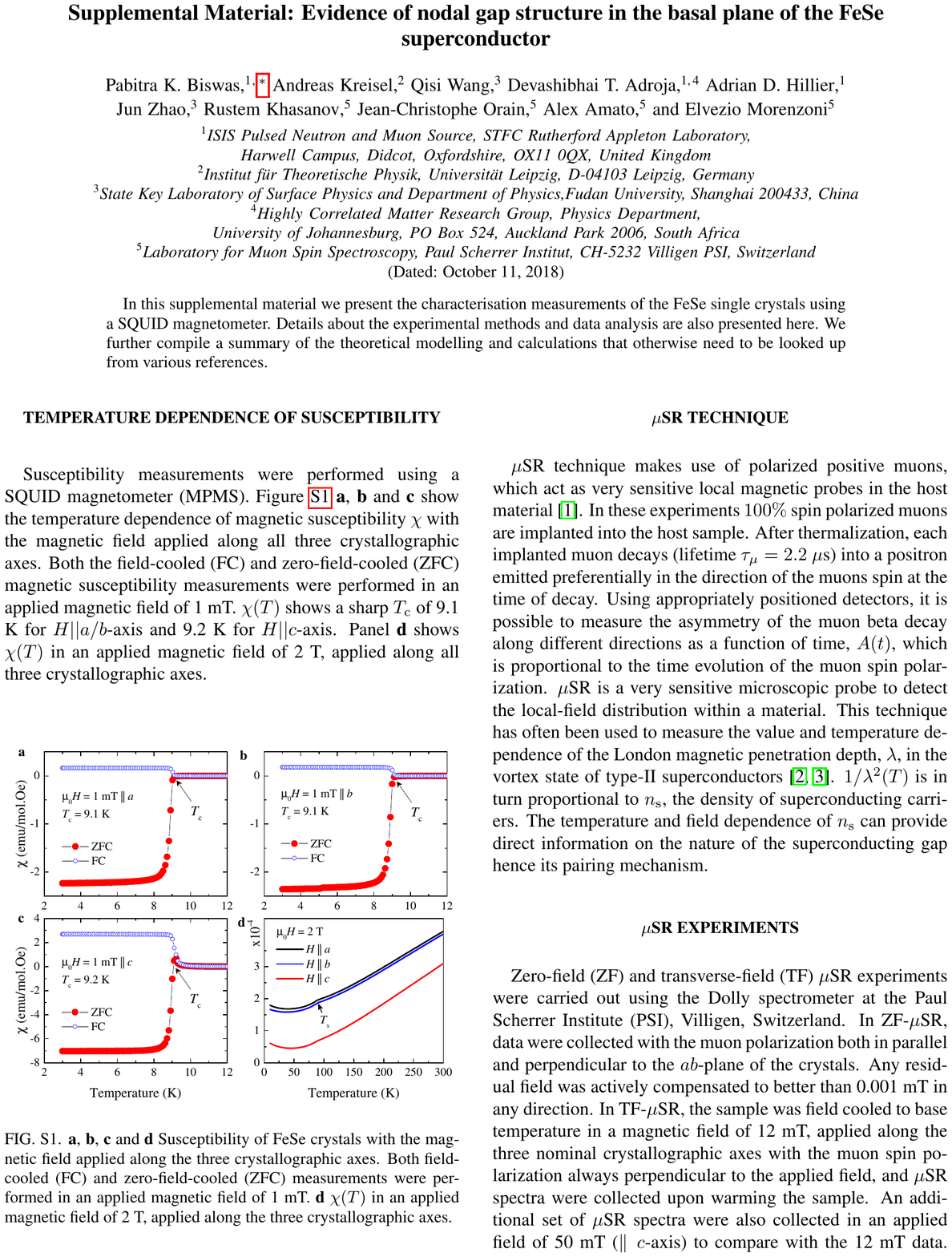}

\pdfximage{\supplementfilename}
\def\numbersupplementpages{\the\pdflastximagepages}


\section*{Supplementary material}

# Supplemental Material: Evidence of nodal gap structure in the basal plane of the FeSe superconductor

Pabitra K. Biswas,[1, *] Andreas Kreisel,[2] Qisi Wang,[3] Devashibhai T. Adroja,[1, 4] Adrian D. Hillier,[1] Jun Zhao,[3] Rustem Khasanov,[5] Jean-Christophe Orain,[5] Alex Amato,[5] and Elvezio Morenzoni[5]

[1]*ISIS Pulsed Neutron and Muon Source, STFC Rutherford Appleton Laboratory, Harwell Campus, Didcot, Oxfordshire, OX11 0QX, United Kingdom*
[2]*Institut für Theoretische Physik, Universität Leipzig, D-04103 Leipzig, Germany*
[3]*State Key Laboratory of Surface Physics and Department of Physics, Fudan University, Shanghai 200433, China*
[4]*Highly Correlated Matter Research Group, Physics Department, University of Johannesburg, PO Box 524, Auckland Park 2006, South Africa*
[5]*Laboratory for Muon Spin Spectroscopy, Paul Scherrer Institut, CH-5232 Villigen PSI, Switzerland*
(Dated: October 11, 2018)

In this supplemental material we present the characterisation measurements of the FeSe single crystals using a SQUID magnetometer. Details about the experimental methods and data analysis are also presented here. We further compile a summary of the theoretical modelling and calculations that otherwise need to be looked up from various references.

## TEMPERATURE DEPENDENCE OF SUSCEPTIBILITY

Susceptibility measurements were performed using a SQUID magnetometer (MPMS). Figure S1 **a**, **b** and **c** show the temperature dependence of magnetic susceptibility $\chi$ with the magnetic field applied along all three crystallographic axes. Both the field-cooled (FC) and zero-field-cooled (ZFC) magnetic susceptibility measurements were performed in an applied magnetic field of 1 mT. $\chi(T)$ shows a sharp $T_c$ of 9.1 K for $H||a/b$-axis and 9.2 K for $H||c$-axis. Panel **d** shows $\chi(T)$ in an applied magnetic field of 2 T, applied along all three crystallographic axes.

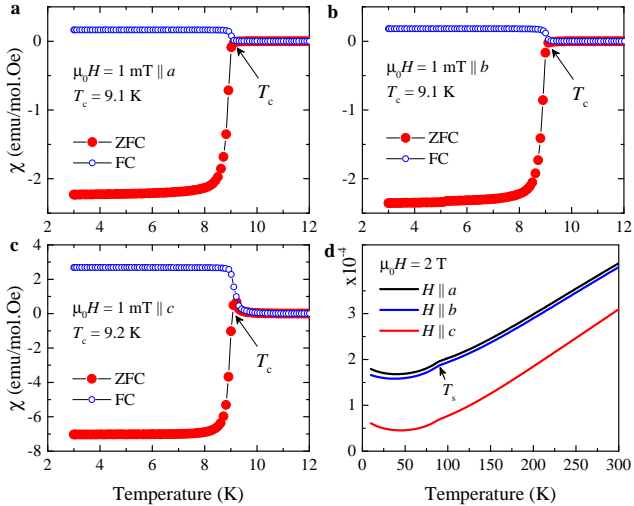

FIG. S1. **a**, **b**, **c** and **d** Susceptibility of FeSe crystals with the magnetic field applied along the three crystallographic axes. Both field-cooled (FC) and zero-field-cooled (ZFC) measurements were performed in an applied magnetic field of 1 mT. **d** $\chi(T)$ in an applied magnetic field of 2 T, applied along the three crystallographic axes.

## $\mu$SR TECHNIQUE

$\mu$SR technique makes use of polarized positive muons, which act as very sensitive local magnetic probes in the host material [1]. In these experiments 100% spin polarized muons are implanted into the host sample. After thermalization, each implanted muon decays (lifetime $\tau_\mu = 2.2$ $\mu$s) into a positron emitted preferentially in the direction of the muons spin at the time of decay. Using appropriately positioned detectors, it is possible to measure the asymmetry of the muon beta decay along different directions as a function of time, $A(t)$, which is proportional to the time evolution of the muon spin polarization. $\mu$SR is a very sensitive microscopic probe to detect the local-field distribution within a material. This technique has often been used to measure the value and temperature dependence of the London magnetic penetration depth, $\lambda$, in the vortex state of type-II superconductors [2, 3]. $1/\lambda^2(T)$ is in turn proportional to $n_s$, the density of superconducting carriers. The temperature and field dependence of $n_s$ can provide direct information on the nature of the superconducting gap hence its pairing mechanism.

## $\mu$SR EXPERIMENTS

Zero-field (ZF) and transverse-field (TF) $\mu$SR experiments were carried out using the Dolly spectrometer at the Paul Scherrer Institute (PSI), Villigen, Switzerland. In ZF-$\mu$SR, data were collected with the muon polarization both in parallel and perpendicular to the $ab$-plane of the crystals. Any residual field was actively compensated to better than 0.001 mT in any direction. In TF-$\mu$SR, the sample was field cooled to base temperature in a magnetic field of 12 mT, applied along the three nominal crystallographic axes with the muon spin polarization always perpendicular to the applied field, and $\mu$SR spectra were collected upon warming the sample. An additional set of $\mu$SR spectra were also collected in an applied field of 50 mT ($\parallel$ $c$-axis) to compare with the 12 mT data.

The typical counting statistics were $\sim$ 20 million muon decays per data point. The ZF- and TF-$\mu$SR data were analyzed using the free software package MUSRFIT [4].

## ANALYSIS OF ZF-$\mu$SR DATA

ZF-$\mu$SR time spectra collected above and below $T_c$ with the muon spin polarization $P_\mu$ aligned both parallel to $a$- and $c$-axis were evaluated using the Kubo-Toyabe relaxation function [5] multiplied by an exponential decay,

$$A_{ZF}(t) = A_0 \exp(-\Lambda t)$$
$$+ \frac{A_{Cu}}{3}\left\{1 + 2\left(1 - \sigma_{Cu}^2 t^2\right)\exp\left(-\frac{\sigma_{Cu}^2 t^2}{2}\right)\right\}, \quad (S1)$$

where $A_0$ and $A_{Cu}$ are the initial asymmetries of the sample and background (due to some muons stopping in the copper foil) signals, respectively, $\sigma_{Cu}$ is the muon spin relaxation rate of the Cu nuclear moments, and $\Lambda$ is the muon spin relaxation rate of the electronic moments present in FeSe. Since we expect that the contribution from the Cu nuclear moments will be similar above and below $T_c$, $\sigma_{Cu}$ was kept as a common parameter for both pair of data sets. For $P_\mu \parallel a-$axis, the fits yield $A_0 = 0.169(6)$, $A_{Cu} = 0.024(6)$, $\sigma_{Cu} = 0.38(2)$ $\mu$s$^{-1}$, $\Lambda(12K) = 0.015(12)$, and $\Lambda(2K) = 0.016(12)$. For $P_\mu \parallel c-$axis, the fits yield $A_0 = 0.217(7)$, $A_{Cu} = 0.027(6)$, $\sigma_{Cu} = 0.35(2)$ $\mu$s$^{-1}$, $\Lambda(14K) = 0.010(6)$, and $\Lambda(2K) = 0.009(6)$. $\sigma_{Cu}$ are relatively large in both sets of data which indicate that some of the muons with lower energy are indeed stopping in the copper foil of the sample holder. The values of $\Lambda$ are also very similar for the data collected above and below $T_c$ in both directions, indicating absence of any detectable magnetic anomaly in the superconducting state of FeSe along both crystallographic directions. The small values of $\Lambda$ are consistent with the presence of diluted and randomly oriented electronic moments in this material.

## ANALYSIS OF TF-$\mu$SR DATA

TF-$\mu$SR asymmetry spectra collected for $H$ applied parallel to the $a$-, $b$-axis were analyzed using an oscillatory term with a Gaussian decay envelope,

$$A_{TF}(t) = A_0 \exp\left(-\sigma^2 t^2/2\right)\cos\left(\gamma_\mu \langle B\rangle t + \phi\right)$$
$$+ A_{bg}\cos\left(\gamma_\mu B_{bg} t + \phi\right), \quad (S2)$$

where $A_0$ and $A_{bg}$ are the initial asymmetries of the sample and background signals, respectively, $\gamma_\mu/2\pi = 135.5$ MHz/T is the muon gyromagnetic ratio [2], $\langle B\rangle$ and $B_{bg}$ are the internal and background magnetic fields, $\phi$ is the initial phase of the muon precession signal, and $\sigma$ is the Gaussian muon spin relaxation rate representing the second moment of the internal field distribution.

In order to account for the highly asymmetric nature of $p(B)$, TF-$\mu$SR asymmetry spectra collected for $H$ applied parallel to the $c$-axis were analyzed using the skewed Gaussian (SKG) field distribution, defined as

$$p_{SKG}(B) = \frac{\sqrt{2/\pi}\gamma}{\sigma_+ + \sigma_-}\begin{cases}\exp\left[-\frac{1}{2}\frac{(B-B_0)^2}{(\sigma_+/\gamma)^2}\right], & B \geq B_0 \\ \exp\left[-\frac{1}{2}\frac{(B-B_0)^2}{(\sigma_-/\gamma)^2}\right], & B < B_0\end{cases}$$
(S3)

where $B_0$ is the field corresponding to the peak value of $p_{SKG}(B)$, $\sigma_+$ and $\sigma_+$ are the Gaussian widths of the SKG field distribution above and below $B_0$, respectively. The first and second moments of $p_{SKG}(B)$ can be written as

$$\langle B\rangle = B_0 + \sqrt{\frac{2}{\pi}}\frac{\sigma_+ - \sigma_-}{\gamma} \quad (S4)$$

and

$$\langle \Delta B^2\rangle = \frac{\sigma_{sc}^2}{\gamma_\mu^2} = \frac{(\pi-2)\sigma_-^2 - (\pi-4)\sigma_+\sigma_- + (\pi-2)\sigma_+^2}{\pi\gamma^2}.$$
(S5)

TF-$\mu$SR asymmetry time spectra were fitted by transforming from the field domain to the time domain via

$$P_{SKG}(t) = \int_{-\infty}^{\infty} p_{SKG}(B)\cos\left(\gamma_\mu B t\right) dB. \quad (S6)$$

TF-$\mu$SR asymmetry spectra were fitted in two steps. First the data were fitted at each temperature with $A_0$, $A_{bg}$, $\langle B\rangle$, $B_{bg}$ and $\sigma$ as common variables. The fits were checked over the entire temperature range to ensure that physical values were obtained for all the parameters at each temperature point. As expected, we found the values of $A_0$ and $A_{bg}$ are mostly temperature independent. To ensure stability of the fits, averaged values of $A_0$ and $A_{bg}$ were then used to refit the data at each temperature point. We obtained $A_0 = 0.078(1)$ and $A_{bg} = 0.116(2)$ for $H$ applied parallel to the $a$-, $b$-axis. For $H$ applied parallel to the $c$-axis, $A_0 = 0.161(1)$ and $A_{bg} = 0.032(1)$.

We have also analyzed the data using the standard 1-component Gaussian equation, often used for powder samples (an oscillatory term with a Gaussian decay envelope, Eq. S2 in the text). While the quality of the fit to the data is worse for this model, we find the temperature dependence of sigma (shown below) very similar to the one extracted from our skewed Gaussian field distribution model.

In our third attempt, we have determined the second moment of the magnetic field distribution by fitting the muon time spectra using a sum of $N = 3$ Gaussian components: [6]

$$A(t) = \sum_{i=1}^{N} A_i \exp(-\sigma_i^2 t^2/2)\cos(\gamma_\mu B_i t + \phi)$$
$$+ A_{bg}\cos(\gamma_\mu B_{bg} t + \phi), \quad (S7)$$

where $\phi$, $A_i$, $\sigma_i$, and $B_i$ are the initial phase, asymmetry, relaxation rate, and mean field (first moment) of the $i$th Gaussian component, respectively. $A_{bg}$ and $B_{bg}$ are the asymmetry and field, respectively due to background contribution, mainly originating from the muons that miss the sample and hit the Cu sample holder. For $N = 3$, the first and second moments of $p(B)$ are given by

$$\langle B \rangle = \sum_{i=1}^{3} \frac{A_i B_i}{A_1 + A_2 + A_3}, \quad (S8)$$

and

$$\langle \Delta B^2 \rangle = \sum_{i=1}^{3} \frac{A_i}{A_1 + A_2 + A_2} \left\{ (\sigma_i/\gamma_\mu)^2 + [B_i - \langle B \rangle]^2 \right\}, \quad (S9)$$

Figure S3 shows the temperature dependence of $\sigma_{ab}$, extracted from the 3-component Gaussian model fit. $\sigma_{ab}$ extracted from all three different models show very similar temperature dependency which proves that independent of the model we use to analyze the TF-$\mu$SR data, $\sigma_{ab}(T)$ and hence $\lambda^{-2}(T)$ are not affected. Only the absolute value of $\lambda(T)$ changes slightly which will not change our main conclusion of this work, i.e. the observation of nodal superconductivity in the basal ($ab-$) plane of FeSe superconductor.

## ANALYSIS OF $\lambda_{ab}$ AND $\lambda_c$

Within the Ginzburg-Landau theory of the vortex state, E. H. Brandt [3] has shown that in extreme type-II superconductor (which is the case for FeSe), $\sigma_{sc}$ is related to the penetra-

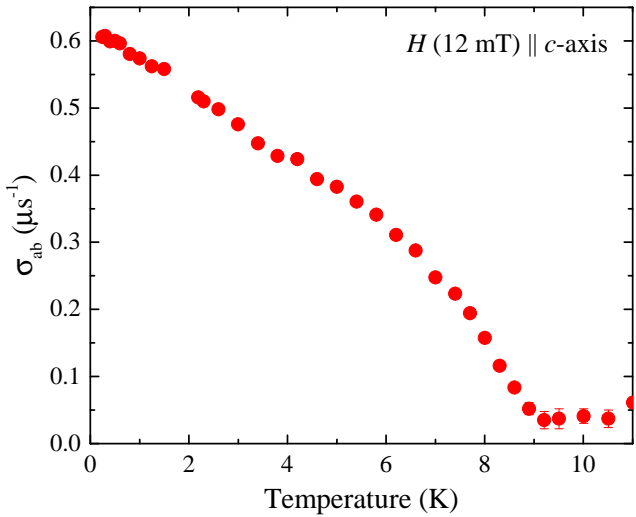

FIG. S2. Temperature dependence of $\sigma_{ab}$, extracted from the standard 1-component Gaussian model fit.

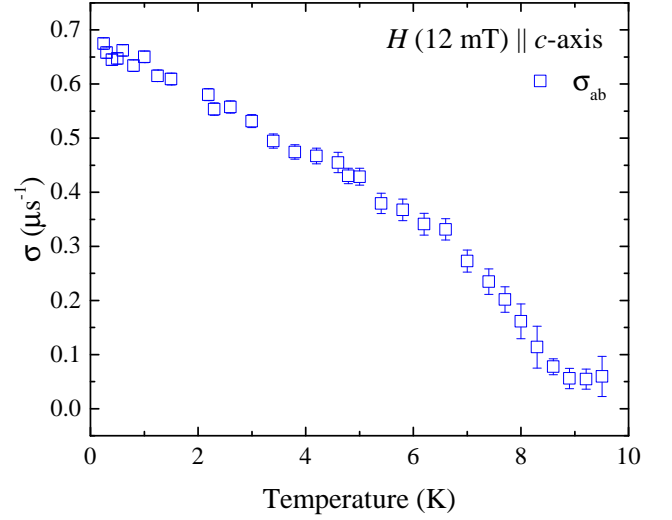

FIG. S3. Temperature dependence of $\sigma_{ab}$, extracted from the 3-component Gaussian model fit.

tion depth $\lambda$ by the simplified equation

$$\frac{\sigma_{sc}(T)}{\gamma_\mu} = 0.06091 \frac{\Phi_0}{\lambda^2(T)}, \quad (S10)$$

where $\Phi_0 = 2.068 \times 10^{-15}$ Wb is the flux quantum. $\lambda^{-2}(T)$ is proportional to the effective superfluid density, $\rho_s \propto \lambda^{-2} \propto n_s/m^*$ ($n_s$ is the charge carrier concentration, and $m^*$ is the effective mass of the charge carriers) and its temperature dependence bear the signature of the symmetry of the superconducting gap. For a highly anisotropic superconductor, the effective penetration depth for the magnetic field applied along the $i^{\text{th}}$ principal axis is then given as [7]

$$\frac{1}{\lambda_{jk}^2} = \frac{1}{\lambda_j \lambda_k} \propto \sigma_{sc}^{\|i}. \quad (S11)$$

This is still true for any anisotropic superconductor such that we can set $\lambda_a = \lambda_b$ for FeSe in the following. The *in-plane* component of the magnetic penetration depth $\lambda_{ab}$ can be obtained from $\sigma_{sc}^{\|c}$ and combining Eq. S10 and Eq. S11 as

$$\frac{1}{\lambda_{ab}^2} = 9.32(\mu m^{-2}/\mu s^{-1}) \times \sigma_{sc}^{\|c}(\mu s^{-1}). \quad (S12)$$

Similarly, the *out-of-plane* component of the magnetic penetration depth $\lambda_c$ can be calculated from $\sigma_{sc}^{\|a}$, $\sigma_{sc}^{\|b}$ and $\sigma_{sc}^{\|c}$ as

$$\frac{1}{\lambda_c^2} = 9.32(\mu m^{-2}/\mu s^{-1}) \times \frac{\sigma_{sc}^{\|a}(\mu s^{-1}) \times \sigma_{sc}^{\|b}(\mu s^{-1})}{\sigma_{sc}^{\|c}(\mu s^{-1})}. \quad (S13)$$

Since, we didn't have all the equivent temperature points in the data sets of $\sigma_{\text{sc}}^{\|a}$ and $\sigma_{\text{sc}}^{\|b}$, for simplicity we have used $\sigma_{\text{sc}}^{\|b}$ in place for $\sigma_{\text{sc}}^{\|a}$ in Eq. S13. This is valid here as the temperature dependence of $\sigma_{\text{sc}}^{\|a}$ and $\sigma_{\text{sc}}^{\|b}$ are identical.

## CALCULATION OF THE PENETRATION DEPTH FROM A TIGHT-BINDING MODEL

### Model of the electronic structure and pairing in FeSe

For the theoretical calculations presented in this section, we start from a multi-band Hamiltonian given by the tight-binding model[8–10]

$$H = \sum_{\mathbf{k}\sigma\ell\ell'} t_{\mathbf{k}}^{\ell\ell'} c_{\ell\sigma}^\dagger(\mathbf{k}) c_{\ell'\sigma}(\mathbf{k}), \quad \text{(S14)}$$

where $c_{\ell\sigma}^\dagger(\mathbf{k})$ is the Fourier amplitude of an operator that creates an electron in Wannier orbital $\ell$ with spin $\sigma$ and $t_{\mathbf{k}}^{\ell\ell'}$ is the Fourier transform of the hopping. Next, we use a modified spin-fluctuation theory that takes into account the reduced coherence of electronic states in certain orbitals[9]. The employed parametrization of the Green's function in terms of quasiparticle weights is

$$\tilde{G}_{\ell\ell'}(\mathbf{k}, \omega_n) = \sqrt{Z_\ell Z_{\ell'}} \sum_\mu \frac{a_\mu^\ell(\mathbf{k}) a_\mu^{\ell'*}(\mathbf{k})}{i\omega_n - \tilde{E}_\mu(\mathbf{k})}, \quad \text{(S15)}$$

where $\tilde{E}_\mu(\mathbf{k})$ are the renormalized band energies and $a_\mu^\ell(\mathbf{k})$ are the matrix elements of the unitary transformation from orbital to band space. Application of this transformation makes the Hamiltonian $H$ diagonal $H = \sum_{\mathbf{k}\sigma\mu} \tilde{E}_\mu(\mathbf{k}) c_{\mu\sigma}^\dagger(\mathbf{k}) c_{\mu\sigma}(\mathbf{k})$ with the true eigenenergies $\tilde{E}_\mu(\mathbf{k})$. In Fig. S4(a) we show the Fermi surface for FeSe derived from this model which are corrugated tubes identified as $\alpha$, $\delta$ and $\varepsilon$ sheets.

For the following, we use a gap function $\Delta_\mathbf{k} = \Delta_0 g(\mathbf{k})$ with a suitable prefactor $\Delta_0(T)$ and a function $g(\mathbf{k})$ that exhibits nodes on the $\varepsilon$ sheet following the behavior evidenced by the bulk $\mu$SR measurements as presented in the main text. Note that this small change from nodeless to nodal can be easily understood in terms of slightly different magnitudes of the angular harmonics that are usually referred to as sign changing s-wave and d-wave in the tetragonal system.

### Penetration depth from a tight-binding approach

For the calculation of the penetration depth that reflects the properties of low-energy excitations in the system, we follow[11, 12]. From the current-current correlator together with the parametrization of the Green's function, we obtain

$$\frac{1}{\lambda_i^2} = \frac{4\pi e^2}{c^2 \hbar^2} \sum_{\mathbf{k},\nu} \frac{d\tilde{E}_\nu(\mathbf{k})}{dk_i} \left( \frac{d\tilde{E}_\nu(\mathbf{k})}{dk_i} |\Delta_\mathbf{k}|^2 - \frac{d|\Delta_\mathbf{k}|}{dk_i} |\Delta_\mathbf{k}| \tilde{E}_\nu(\mathbf{k}) \right)$$
$$\times \frac{\tilde{Z}_\nu(\mathbf{k})}{E_{\nu,\mathbf{k}}^2} \left( \frac{1}{E_{\nu,\mathbf{k}}} \tanh\left(\frac{E_{\nu,\mathbf{k}}}{2k_B T}\right) - \frac{1}{2k_B T} \text{sech}\left(\frac{E_{\nu,\mathbf{k}}}{2k_B T}\right)^2 \right). \quad \text{(S16)}$$

where $E_{\nu,\mathbf{k}} = \sqrt{\tilde{E}_\nu(\mathbf{k})^2 + |\Delta_\mathbf{k}|^2}$ are the Bogoliubov quasiparticle energies and $\tilde{Z}_\nu(\mathbf{k}) = (\sum_\ell |a_\nu^\ell(\mathbf{k})|^2 \sqrt{Z_s})^2$ are the quasiparticle weights of band $\nu$ near the Fermi surface. For our calculation, we use the superconducting gap function $\Delta_\mathbf{k}$ as discussed above with a mean field like $T$ dependence of the order parameter $\Delta_\mathbf{k} = g(\mathbf{k}) \Delta_0 \tanh(1.76 \cdot \sqrt{T_c/T - 1})$[12]. The momentum sum is evaluated for $\approx 10^6$ $k$-points to obtain the values of the penetration depth tensor along the 3 principal directions $x$, $y$, $c$, where the first two are along the Fe-Fe bond directions[10] and the third along the crystallographic axis out of plane. Note that for $i = x$, the region of small gap on the $\epsilon$ pocket will show up in the behavior of the penetration depth at low temperatures, while for $i = y$, the region of small gap on the $\alpha$ pocket determines the small temperature properties of the penetration depth. Since the $\delta$ pocket is not seen in spectroscopic probes, we present results where the contribution of this Fermi surface sheet is not taken into account; the differences to the full calculations are small because the quasiparticle weight $\tilde{Z}_\nu(\mathbf{k})$ is small anyhow[9].

### Discussion of results and connection to experimental investigations

As outlined from the previous paragraph, all relevant quantities to calculate the penetration depth are already fixed by other experiments, such that there is in principle no free parameter within this theoretical model. One exception might however be the influence of the $\delta$ pocket that contributes in the calculation, but has not been observed with spectroscopic probes so far. In Figure S4(b), the result of the evaluation of Eq. (S16) is shown by excluding the contribution of the $\delta$ pocket (full lines), together with a calculation where also the $\delta$ pocket is taken into account (dashed lines). In both quantities (with and without $\delta$ pocket), it can be observed that $1/\lambda^2$ has the same order of magnitude for the $x$ and $y$ directions, but is much smaller for the $c$ direction (not shown). Considering the model for the electronic structure, this is expected and can be read off from Eq. (S16). Noting that there are not qualitative differences in the behavior of $1/\lambda^2$ for the calculation with and without contributions from the $\delta$ pocket, and considering that there is (to our knowledge) no experimental data available on the gap structure of this pocket, we decide to not discuss the influence of the $\delta$ pocket further. Looking at the absolute numbers, it seems that the calculation without $\delta$ pocket agrees better with the measured $1/\lambda^2$ pointing towards that it does not contribute to superconductivity as also

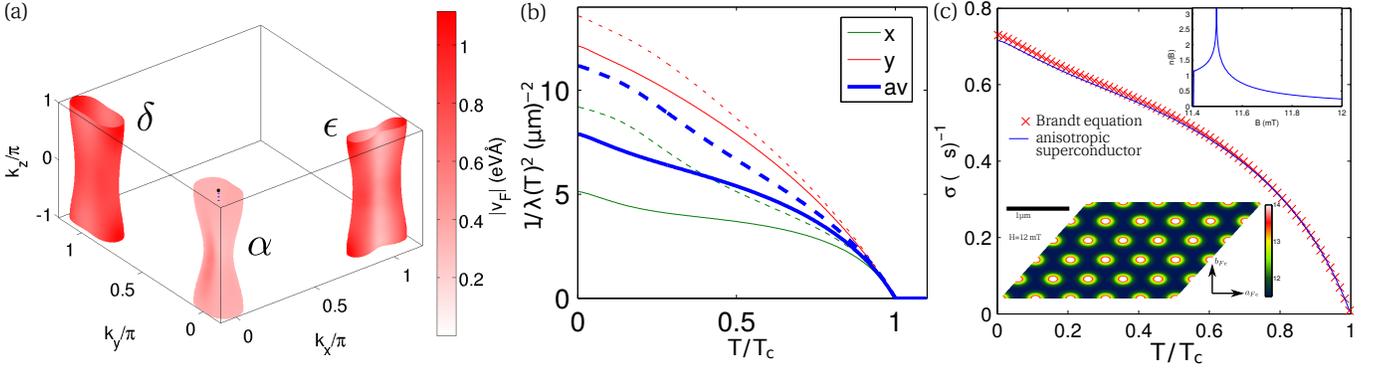

FIG. S4. (a) Fermi surface of the model used for the microscopic calculations with the 3 Fermi surface pockets $\alpha$, $\epsilon$ and $\delta$, the Fermi velocity is plotted color-coded. (b) Calculated penetration depth for shielding currents along the directions in the basal plane, and its geometric average according to Eq. (S11). The full lines are the calculation when ignoring the contribution from the $\delta$ pocket and the dashed lines are from the calculation where this constraint is removed. (c) Calculated broadening using the second moment of simulated field distributions (solid line) compared to the result from the Brandt equation using the geometric average, corresponding experimental result is shown in Fig. 2 **i** of the main text. Upper inset: Example of such a field distribution at $T = 0$, lower inset: magnitude of the field plotted in real space.

TABLE I. Fitted parameters to the $\lambda_{ab}^{-2}(T)$ data of FeSe (for $H = 50$ mT $||c$-axis) using the different models as described in the text.

| Data | Model | Gap value (meV) | $\lambda(0)(nm)$ | $\chi^2_{reduced}$ |
|---|---|---|---|---|
| | s wave | $\Delta$=1.27(2) | | 18.6 |
| | Anisotropic s wave | $\Delta$=1.41(3), $a$=0.73(3) with $\Delta_{Max}$=2.43(5) | | 2.5 |
| $\frac{1}{\lambda_{ab}^2(T)}$ | d wave | $\Delta$=2.03(4) | | 2.4 |
| | $s + s$ wave | $\Delta_1$=1.9(1), $\Delta_2$=0.50(6) and $\omega$=0.66(3) | | 2.6 |
| | $s + d$ wave | $\Delta_1$=1.8(1), $\Delta_2$=0.8(2) and $\omega$=0.61(4) | 413(31) | 1.8 |

proposed theoretically by other approaches[13, 14]. The second term (derivative of order parameter with respect to the momentum parallel to the direction of the penetration depth) does not contribute significantly to the final result. Thus, the sum is dominated by terms where the gap $\Delta_\mathbf{k}$ and the projection of the Fermi velocity $\frac{d\tilde{E}_\nu(\mathbf{k})}{dk_i}$ are large. This result is well known, $\lambda_i^{-2} \propto \langle \Delta_\mathbf{k} v_i^2 \rangle$ and the deviations between theoretical result $\lambda_{av}$ (Fig. 3 **d** of the main text) and the experimentally deduced $\lambda_{ab}$ can easily be explained by the uncertainties of the Fermi velocities and/or the gap magnitudes $\Delta_0$, both of them have just been kept identical to those of Refs.[9, 10]. Noting that the system is very two dimensional (only very small dispersion in $k_z$ direction), which is in agreement to expectations from *ab initio* calculations and has been verified experimentally by ARPES measurements[15–18], it is reasonable to assume that the projection of the Fermi energies in $k_z$ direction is small, an assumption that produces $\lambda_c^{-2}$ much too small compared to the experimental result.

Finally, to make connection to experimental results for the $1/\lambda^2$ as obtained from measurements on twinned crystals, we simply calculate the geometric average in the a-b plane $1/\lambda_{av}^2 = 1/(\lambda_x \lambda_y)$ for the two cases discussed above. The correctness of Eq. (S10, S11) and finally also (S12,S13) has been checked by solving the London equation in the vortex state as summarized in the next section.

### Calculation of field distribution in vortex state

Following Ref. [19, 20], a generalized mass tensor $M$ is introduced to write down the London free energy in terms of the magnetic field $\vec{H}(\vec{r})$ inside the superconductor

$$F = \mu_0 \int \left( \vec{H}^2 + \lambda^2 \sum_{i,k=1}^{3} m_{ik} [\vec{\nabla} \times \vec{H}]_i [\vec{\nabla} \times \vec{H}]_k \right) d^3\vec{r}, \quad \text{(S17)}$$

where $\mu_0$ is the Bohr magneton and $m_{ik}$ are the elements of the mass tensor. The mass tensor $M$ is symmetric and can be diagonalized in the crystal frame and is normalized to one $m_{11} m_{22} m_{33} = 1$ such that $\lambda$ is the geometric mean of the penetration depth in the 3 directions. For arbitrary (external) field directions, we can rotate the coordinate system around an arbitrary axis by a rotation matrix $D$ which will transform the mass tensor according to $\tilde{M} = D^T M D$. The anisotropic London equation for this general case in presence of vortices

with flux $\phi_0$ at positions $\vec{r}_\nu$ is then given by

$$H_i = \lambda^2 \sum_{k,l,s,t,j=1}^{3} \tilde{m}_{kl}\epsilon_{lsi}\epsilon_{ktj}\frac{\partial^2 H_j}{\partial x_s \partial x_t} + \delta_{i3}\sum_\nu \phi_0 \delta(\vec{r}-\vec{r}_\nu). \quad (S18)$$

Using $\nabla \cdot H = 0$ and the symmetry $\partial_3 H_i = 0$, one obtains $\partial_1 H_1 = -\partial_2 H_2$ that can be used to simplify the equation above. Assuming a flux line lattice, one can solve the differential equation by Fourier transformation

$$H_k(\vec{r}) = \sum_{\vec{G}} H_k^{\vec{G}} e^{i\vec{G}\cdot\vec{r}} \quad (S19)$$

where the sum runs over the reciprocal lattice vectors of the vortex lattice. The resulting algebraic equation can be written as

$$A\vec{H}^{\vec{G}} = \vec{C} \quad (S20)$$

with $\vec{C} = (0, 0, \phi_0)$ and a matrix

$$A = \begin{pmatrix} 1+n_{33}G^2 & 0 & -n_{13}G_2^2 + n_{23}G_1G_2 \\ 0 & 1+n_{33}G^2 & n_{13}G_1G_2 - n_{23}G_1^2 \\ -n_{31}G^2 & -n_{23}G^2 & 1+n_{11}G_2^2+n_{22}G_1^2-2n_{21}G_1G_2 \end{pmatrix}, \quad (S21)$$

where $G^2 = G_1^2 + G_2^2$ and $n_{ik} = \lambda^2 \tilde{m}_{ik}$. This equation has the solution $\vec{H}^{\vec{G}} = A^{-1}\vec{C}$ such that the field in real space can be calculated by performing the lattice sum in Eq. (S19). We assume a distorted hexagonal lattice that is parametrized by two parameters [20] which are determined to minimize the free energy Eq. (S17). The result of such a simulation is shown as lower inset in Fig. (S4) c such that the field distribution can then be calculated efficiently with a two dimensional version of the tetrahedron method resulting in the field distribution as shown in the upper inset of Fig. (S4) c. The resulting field distribution is used to calculate the second moment $\langle \Delta B^2 \rangle$ and directly simulate $\sigma_c$ for the anisotropic vortex state. In Fig. S4 **C**, we show the result of this approach in comparison with the expected result from Eq. (S10, S11). Those are in excellent agreement, proving that the use of the latter equations is suitable to analyze also fully anisotropic superconductors. We furthermore checked the influence of a small misalignment of the field for measurements where it is in the basal plane and found that this in principle slightly increases the broadening $\sigma_a$ because components of the mass tensor in the plane are picked up, but cannot explain the large experimental broadening as compared to the theoretical expectations. In summary, the requirement of the larger dispersion/Fermi velocities in $k_z$ direction is unchanged upon this analysis.

## $\lambda_{ab}(T)$ FOR $H = 50$ MT $\parallel$ $c$-AXIS

We have performed all the TF-$\mu$SR measurements in a magnetic field of $H = 12$ mT. This is due to the limitation of the field range that can be applied along $a$- and $b$-axis. However

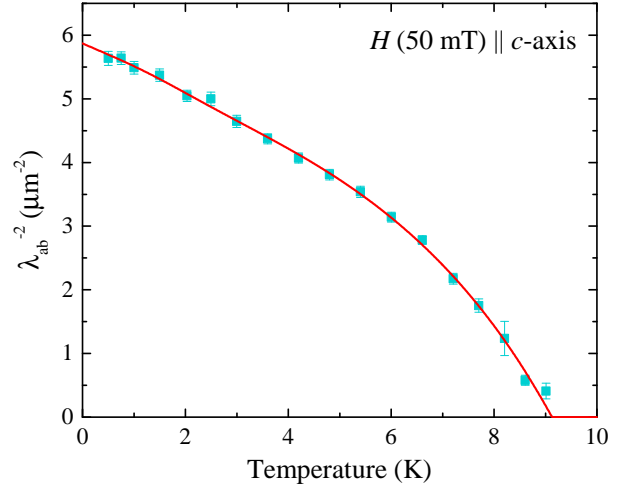

FIG. S5. Temperature dependence of $\lambda_{ab}^{-2}$ for FeSe with a magnetic field $H = 50$ mT applied along the $c$-axis. The solid curve is the fit to the $\lambda_{ab}^{-2}(T)$ data using the two-gap $s + d$ wave model.

there is no such limitation of the applied field along $c$-axis. Therefore, we have collected another set of TF-$\mu$SR data at a higher field of $H = 50$ mT applied parallel to the $c$-axis to compare with the data collected at $H = 12$ mT. Figure S5 shows the temperature dependence of $\lambda_{ab}^{-2}$ for FeSe with the field $H = 50$ mT applied along the $c$-axis. The solid curve is the fit to the $\lambda^{-2}(T)$ data using a two-gap $s + d$ wave model. All the fitted parameters are summarized in Table I. Here again we find that $s + d$ wave model gives much lower $\chi^2_{\text{reduced}}$ value than any other models. All the fitted parameters for this set of data are consistent with data data collected at $H = 12$ mT which further suggest the presence of a nodal gap in the basal plane of FeSe superconductor. This also proves that an applied field of 12 mT was sufficient enough to produce stable vortices in the superconducting state of FeSe. The estimated $n_s^{\parallel ab}(0) \approx 6.6 \times 10^{20}$ cm$^{-3}$ is again consistent with the previous value.

---


\end{document}